\documentclass[a4paper,12pt]{article}
\usepackage{amssymb}
\usepackage{amsmath}
\usepackage{fullpage}
\usepackage{boxedminipage}
\usepackage{listings}
\usepackage{minitoc}

\makeatletter \@addtoreset{equation}{section} \makeatother
\renewcommand{\theequation}{\thesection.\arabic{equation}}

\begin{document}

\begin{titlepage}

    \thispagestyle{empty}
    \begin{flushright}
        \hfill{CERN-PH-TH/164} \\\hfill{UCLA/07/TEP/19}\\
    \end{flushright}

    \vspace{5pt}
    \begin{center}
        { \LARGE{\textbf{Non-BPS Attractors\\\vskip.5cm
        in 5d and 6d Extended Supergravity}}}\vspace{15pt}
        \vspace{35pt}

        { {\textbf{L.Andrianopoli$^{\heartsuit\diamondsuit\clubsuit}$, S.Ferrara$^{\diamondsuit\spadesuit\flat}$, A.Marrani$^{\heartsuit\spadesuit}$ and M.Trigiante$^{\clubsuit}$}}}\vspace{15pt}

        {$\heartsuit$ \it Museo Storico della Fisica e\\
        Centro Studi e Ricerche ``Enrico Fermi"\\
        Via Panisperna 89A, 00184 Roma, Italy}

        \vspace{10pt}

        {$\diamondsuit$ \it Physics Department,Theory Unit, CERN, \\
        CH 1211, Geneva 23, Switzerland\\
        \texttt{Laura.Andrianopoli@cern.ch, sergio.ferrara@cern.ch}}

        \vspace{10pt}

        {$\clubsuit$ \it Dipartimento di Fisica, Politecnico di Torino,\\
        Corso Duca degli Abruzzi 24, I-10129 Torino, Italy\\
        and INFN - sezione di Torino, Italy\\
        \texttt{mario.trigiante@polito.it}}

        \vspace{10pt}

        {$\spadesuit$ \it INFN - Laboratori Nazionali di Frascati, \\
        Via Enrico Fermi 40,00044 Frascati, Italy\\
        \texttt{marrani@lnf.infn.it}}

        \vspace{10pt}

         {$\flat$ \it Department of Physics and Astronomy,\\
        University of California, Los Angeles, CA USA\\
        \texttt{ferrara@physics.ucla.edu}}

        \vspace{15pt}
\end{center}

\vfill
\begin{abstract}
We connect the attractor equations of a certain class of $N=2$,
$d=5$ supergravities with their $(1,0)$, $d=6$ counterparts, by
relating the moduli space of non-BPS $d=5$ black hole/black string
attractors to the moduli space of extremal dyonic black string $d=6$
non-BPS attractors. For $d = 5$ real special symmetric spaces and
for $N = 4,6,8$ theories, we explicitly compute the flat directions
of the black object potential corresponding to vanishing eigenvalues
of its Hessian matrix. In the case $N = 4$, we study the relation to
the $(2,0)$, $d=6$ theory. We finally describe the embedding of the
$N=2$, $d=5$ magic models in $N=8$, $d=5$ supergravity as well as
the interconnection among the corresponding charge orbits.
\end{abstract}

\end{titlepage}
\newpage

\section{\label{Intro}Introduction}

Recently the study of the attractor equations for extremal black
holes (BHs) \cite{FKS}--\nocite{Strom,FK1,FK2}\cite{FGK} in four
dimensions received special attention, especially in relation with
new results on non-BPS, non-supersymmetric solutions
\cite{Sen-old1}--\nocite
{GIJT,Sen-old2,K1,TT,G,GJMT,Ebra1,K2,Ira1,Tom,BFM,AoB-book,FKlast,Ebra2,%
BFGM1,rotating-attr,K3,Misra1,Lust2,BFMY,CdWMa,DFT07-1,BFM-SIGRAV06,Cer-Dal,%
ADFT-2,Saraikin-Vafa-1,Ferrara-Marrani-1,TT2,ADOT-1,fm07,CCDOP,Misra2,Astefanesei,%
Anber,Myung1,CFM1,BMOS-1,Hotta,Gao,PASCOS07,Sen-review}\cite{Belhaj1}.

Not much is known about non-BPS attractors in five dimensions,
although general results for symmetric special geometries in BHs
(and black strings) backgrounds were derived in \cite{FG2}. More
recently, it has been shown \cite{fm07} that real special symmetric
spaces have, in the non-BPS case, a moduli space of vacua, as it was
the case for their $d=4$ special K\"{a}hler descendants
\cite{BFGM1}. In four dimensions, massless Hessian modes for generic
cubic geometries were shown to occur for the non-BPS case with
non-vanishing central charge in \cite{TT,TT2}. Some additional
insight on the correspondence among (the supersymmetry preserving
features of) extremal BH attractors in four and five dimensions have
been gained in \cite{CFM1}, by relating the $d=4$ and $5$ BH
potentials and the corresponding attractor equations. In particular,
it was shown that the moduli space of non-BPS attractors in $d=5$
real special symmetric
geometries must be in the intersection of the moduli spaces of non-BPS $%
Z\neq 0$ and non-BPS $Z=0$ attractors in the corresponding $d=4$ special
K\"{a}hler homogeneous geometries.

Aim of the present investigation is to perform concrete computations of the
massless modes of the non-BPS $d=5$ Hessian matrix, and further relate the $%
d=5$ BH (or black string) potential to the $d=6$ dyonic extremal black
string potential and its BPS and non-BPS critical points, following the
approach of \cite{FG2} and \cite{ADFL-6}. This analysis reveals a noteworthy
feature of the relation between $d=5$ and $d=6$. Namely, the moduli space of
$d=6$ non-BPS (with vanishing central charge\footnote{%
This means that non-BPS dyonic strings are neutral with respect to the
central extension of the $(1,0)$, $d=6$ supersymmetry algebra.%
}) dyonic string attractors is a submanifold of the moduli space of $d=5$
non-BPS attractors of symmetric real special geometries. The only exception
is provided by the cubic reducible sequence of real special geometries, for
which the non-BPS $d=6$ and $d=5$ moduli spaces actually coincide. It is
worth pointing out that moduli spaces also exist, for particular
non-BPS-supporting charge configurations, for all real special geometries
with a $d=6$ uplift \cite{AFL-6-5}. This is the case for the homogeneous
non-symmetric real special geometries studied in \cite{vanderseypen}. For $%
N=2$, $d=5$ magic supergravities, with the exception of the octonionic case,
the non-BPS moduli spaces can also be obtained as suitable truncations of
the moduli space of BPS sttractors of $N=8$, $d=5$ supergravity. In all
cases, the Hessian matrix is semi-positive definite.

It is worth pointing out that in this work we consider only extremal
black $p $-extended objects which are asymptotically flat,
spherically symmetric and with an horizon geometry $AdS_{p+2}\times
S^{d-p-2}$ \cite{Gibbons-Townsend}. Thus, we do not deal with, for
instance, black rings and rotating BHs in $d=5$, which however also
exhibit an attractor behaviour (see \textit{e.g.} \cite
{Larsen-review}).

The paper is organized as follows.

In Sect. \ref{Sect2} we recall some relevant facts about $N=2$,
$d=6$ self-dual black string attractors and the properties of the
black string effective potential in terms of the moduli space
spanned by the tensor multiplets' scalars. In Sect. \ref{Sect3} we
discuss the $d=5$ effective potential in a six-dimensional language
for the $d=5$ models admitting a $d=6 $ uplift (including all
homogeneous real special geometries classified in
\cite{vanderseypen}), in the absence (Subsect. \ref{Subsect31}) or
presence (Subsect. \ref{Subsect32}) of $d=6$ vector multiplets. In
Subsubsect. \ref {Subsubsect321} we perform an analysis of the
attractors in $d=5$, $N=2$ magic supergravities, and comment on the
moduli spaces of attractor solutions for such theories. Thence, in
Subsects. \ref{Subsect41}, \ref {Subsect42} and \ref{N=4} we recall
a similar analysis of the attractors respectively in $d=5$, $N=8$,
$6$ and $4$ supegravities \cite {FG2,adf5d,ADF-U-duality}. The
analysis holds for all $N=2$ symmetric spaces, as well as for
homogeneous spaces by considering particular charge configurations.
In Sect. \ref{Anomalies} we comment on the conditions to be
satisfied in order to obtain an anomaly-free $(1,0)$, $d=6$
supergravity by uplifting $N=2$, $d=5$ theories. Sect.
\ref{Conclusion} is devoted to final remarks and conclusions.

The Appendix discusses some group embeddings, relevant in order to elucidate
the relation between the $N=8$, $d=5$ BPS unique orbit and the non-BPS
orbits of the $N=2$, $d=5$ theories obtained as consistent truncations of $%
N=8$ supergravity. Such $N=2$ theories include the magic supergravities
based on the Jordan algebras $J_{3}^{\mathbb{H}}$, $J_{3}^{\mathbb{C}}$, $%
J_{3}^{\mathbb{R}}$ with $n_{H}=0,1,2$ hypermultiplets,
respectively. \setcounter{equation}0
\def\theequation{\arabic{section}.\arabic{equation}}

\section{\label{Sect2}$(1,0)$, $d=6$ attractors for extremal dyonic strings}

In $d=6$, $(1,0)$ and $\left( 2,0\right) $ chiral
supergravities\footnote{In the literature they are sometimes
referred to as $(2,0)$ and $(4,0)$ respectively \cite{adf96}.} there
are no BPS BH states, because the central extension of the
corresponding $d=6$ superalgebras does not contain scalar central
charges \cite{Townsend}. However, there are BPS (dyonic) string
configurations, as allowed from the superalgebra, and extremal black
string BPS attractors exist \cite{ADFL-6,FG2}. Such attractors
preserve $4$ supersymmetries, so they are the $d=6$ analogue of
$d=5$ and $d=4$ $\frac{1}{2}$-BPS extremal BH attractors.
Interestingly enough, extremal black string non-BPS attractors also
exist in such $d=6$ theories \cite{FG2}, as it is the case for
(extremal BH attractors) in $d=5$ and $d=4$. The next sections are
partially devoted to such an issue.

Let us start by recalling the general structure of the minimal
supergravity in $d=6$, the chiral $(1,0)$ theory. The field content
of the minimal theory is:

\begin{itemize}
\item  Gravitational multiplet:
\begin{equation}
(V_{\mu }^{a},\psi _{A\mu },B_{\mu \nu }^{+})\,;\qquad ;(\mu =0,1,\cdots
,5;\;A=1,2)\,;
\end{equation}

\item  Tensor multiplets:
\begin{equation}
(B_{\mu \nu }^{-},\chi ^{A},\phi )^{i}\,;\qquad (i=1,\cdots ,q+1)\,;
\end{equation}
The scalar fields in the tensor multiplets sit in the coset space \cite
{Romans}
\begin{equation}
\frac{G}{H}=\frac{O(1,q+1)}{O(q+1)}.
\end{equation}
They may be parametrized in terms of $q+2$ fields $X^{\Lambda }$, ($\Lambda
=0,1,\cdots ,q+1$), contstrained by the relation
\begin{equation}
X^{\Lambda }X^{\Sigma }\eta _{\Lambda \Sigma }\equiv X^{\Lambda }X_{\Lambda
}=1\,,  \label{constr6}
\end{equation}
where $\eta _{\Lambda \Sigma }=diag[1,-1,\cdots ,-1]$. The kinetic
matrix for the tensors is:
\begin{equation}
G_{\Lambda \Sigma }=2X_{\Lambda }X_{\Sigma }-\eta _{\Lambda \Sigma }
\label{gls}
\end{equation}
whose inverse matrix is:
\begin{equation}
G^{\Lambda \Sigma }=2X^{\Lambda }X^{\Sigma }-\eta ^{\Lambda \Sigma }\,.
\end{equation}
As for any $d=6$ theory, the field strengths of the antisymmetric tensors $%
H^{\Lambda }=dB^{\Lambda }$ have definite self-duality properties:
\begin{equation}
G_{\Lambda \Sigma }{}^{\star }H^{\Sigma }=\eta _{\Lambda \Sigma }H^{\Sigma
}\,;
\end{equation}
As a consequence, there is no distinction between the associated electric
and magnetic charges
\begin{equation}
e^{\Lambda }=\eta ^{\Lambda \Sigma }e_{\Sigma }=\int_{S^{3}}H^{\Lambda }\,.
\end{equation}

\item  Vector multiplets:
\begin{equation}
(A_{\mu },\lambda _{A})^{\alpha }\,;\qquad (\alpha =1,\cdots ,m)\,;
\end{equation}
The kinetic matrix for the vector field strengths is given in terms of a
given constant matrix $C_{\alpha \beta }^{\Lambda }$ by \cite{Sagnotti}:
\begin{equation}
\mathcal{N}_{\alpha \beta }=X_{\Lambda }C_{\alpha \beta }^{\Lambda }\,.
\end{equation}

\item  Hypermultiplets:
\begin{equation}
(\zeta ^{A},4q)^{\ell }\,;\qquad (\ell =1,\cdots ,p)\,.
\end{equation}
The hypermultiplets do not play any role in the attractor mechanism, and
will not be discussed further here.
\end{itemize}

Since the vector multiplets do not contain scalar fields, the only
contribution to the black string effective potential comes from the tensor
multiplets, and reads \cite{ADFL-6}:
\begin{equation}
V^{(6)}=G^{\Lambda \Sigma }e_{\Lambda }e_{\Sigma }=2(X^{\Lambda }e_{\Lambda
})^{2}-e^{\Lambda }e_{\Lambda }
\end{equation}
or equivalently, in terms of the dressed central and matter charges $%
Z=(X^{\Lambda }e_{\Lambda })$ and $Z_{i}=P_{i\Lambda }e^{\Lambda }$ (where $%
P^{\Lambda \Sigma }$, $P^{\Lambda \Sigma }X_{\Sigma }=0$ is the projector
orthogonal to the central charge):
\begin{equation}
V^{(6)}=Z^{2}+Z_{i}Z^{i}\,.  \label{V6}
\end{equation}
The criticality conditions for the effective black string potential (\ref{V6}%
) reads
\begin{equation}
\partial _{i}V^{(6)}=0\Leftrightarrow ZZ_{i}=0,~\forall i,
\end{equation}
and therefore two different extrema are allowed, the BPS one for $%
Z_{i}=0~\forall i$, and a non-BPS one for $Z=0$, both yielding the following
critical value of $V^{(6)}$:
\begin{equation}
V^{(6)}|_{extr}=|e^{\Lambda }e_{\Lambda }|\,.
\end{equation}

\setcounter{equation}0
\def\theequation{\arabic{section}.\arabic{equation}}

\section{\label{Sect3}$N=2$, $d=5$ attractors with a six dimensional interpretation}

In the absence of gauging, the minimal five dimensional theory generally
admits the following field content (omitting hypermultiplets):

\begin{itemize}
\item  Gravitational multiplet:
\begin{equation}
(V_{\mu }^{a},\psi _{A\mu },A_{\mu })\,;\qquad (\mu =0,1,\cdots
,4;\;A=1,2)\,;
\end{equation}

\item  Vector multiplets:
\begin{equation}
(A_{\mu },\chi ^{A},\phi )^{a}\,;\qquad (a=1,\cdots ,n)\,.
\end{equation}
\end{itemize}

The scalar fields do not necessarily belong to a coset manifold, but their $%
\sigma $-model is described by real-special geometry. In particular, the
scalar manifold is described by the locus
\begin{equation}
\mathcal{V}(L)=1  \label{constr5}
\end{equation}
where $L^{I}(\phi )$, $I=0,1,\cdots ,n$ are function of the scalars and $%
\mathcal{V}$ is the cubic polynomial:
\begin{equation}
\mathcal{V}(L)=\frac{1}{3!}d_{IJK}L^{I}L^{J}L^{K}\,,  \label{polyn}
\end{equation}
written in terms of an appropriate totally symmetric, constant matrix $%
d_{IJK}$. Note that in order to have a $d=6$ uplift the real special
geometry must have a certain structure, as discussed in \cite{FMS}. Namely
\begin{equation}
\mathcal{V}=zX^{\Lambda }\eta _{\Lambda \Sigma }X^{\Sigma }+X^{\Lambda
}C_{\Lambda \alpha \beta }X^{\alpha }X^{\beta } .  \label{FMS1}
\end{equation}
This is always the case for the homogeneous spaces discussed in \cite
{vanderseypen}, where $C_{\Lambda \alpha \beta }$ is written in terms of the
$\gamma $-matrices of $SO\left( q+1\right) $ Clifford algebras.

The kinetic matrix for the vector field-strengths has the general form:
\begin{equation}
a_{IJ}=-\partial _{I}\partial _{J}\log \mathcal{V}|_{\mathcal{V}=1}\,.
\label{aij}
\end{equation}

The BH effective potential in five dimensions is given by
\begin{equation}
V^{(5)}=a^{IJ}q_{I}q_{J}  \label{veff5}
\end{equation}
where $q_{I}=\int_{S^{3}}\frac{\partial \mathcal{L}}{\partial
F^{I}}$ are the electric charges and $a^{IJ}$ the inverse of
(\ref{aij}). \setcounter{equation}0
\def\theequation{3.\arabic{subsection}.\arabic{equation}}

\subsection{\label{Subsect31}No $d=6$ vector multiplets}

We are interested in finding the relation of the six dimensional attractor
behavior to the five dimensional one. Let us first consider the simplest
case of a six dimensional supergravity theory only coupled to $q+1$ tensor
multiplets (no vector multiplets). In this case, $n=q+1$ and the scalar
content is given by the six dimensional scalars $X^{\Lambda }$ plus the
Kaluza--Klein (KK) dilaton $z$. The five dimensional scalar fields are
related by the constraint (\ref{constr5}), where the surface expression (\ref
{polyn}) takes here the simple form\footnote{%
This corresponds to the $d=5$ symmetric real spaces of the
\textit{``}generic sequence'' $SO\left( 1,1\right) \times
\frac{SO\left( 1,q+1\right) }{SO\left( q+1\right) }$ \cite{gst}.}:
\begin{equation}
\mathcal{V}(L)=\mathcal{V}(z,X)=\frac{1}{2}zX^{\Lambda }X^{\Sigma }\eta
_{\Lambda \Sigma }\,.
\end{equation}
The constraint (\ref{constr5}) then becomes:
\begin{equation}
\frac{1}{2}X^{\Lambda }X_{\Lambda }=z^{-1}\,.
\end{equation}

The components of the kinetic matrix are in this case:
\begin{equation}
a_{IJ}=\left\{
\begin{matrix}
a_{zz} & = & z^{-2}\cr a_{z\Lambda } & = & 0\cr a_{\Lambda \Sigma } & = & z\,%
\widetilde{G}_{\Lambda \Sigma }\cr
\end{matrix}
\right.  \label{aij1}
\end{equation}
where the matrix $\widetilde{G}$
\begin{equation}
\widetilde{G}_{\Lambda \Sigma }(X)=2\frac{X_{\Lambda }X_{\Sigma }}{X^{\Gamma
}X_{\Gamma }}-\eta _{\Lambda \Sigma }
\end{equation}
is related to $G$ in (\ref{gls}) by
\begin{equation}
\widetilde{G}_{\Lambda \Sigma }|_{X^{\Lambda }X_{\Lambda }=1}=G_{\Lambda
\Sigma }\,.
\end{equation}
More precisely, setting:
\begin{equation}
\hat{X}^{\Lambda }\equiv \frac{X^{\Lambda }}{\sqrt{X^{\Lambda }X_{\Lambda }}}%
\,,\qquad (\hat{X}^{\Lambda }\hat{X}_{\Lambda }=1)
\end{equation}
we have:
\begin{equation}
\widetilde{G}_{\Lambda \Sigma }(X)=G_{\Lambda \Sigma }(\hat{X})\,.
\label{tildeg}
\end{equation}

The matrix (\ref{aij1}) is easily inverted giving:
\begin{equation}
a^{IJ}=\left\{
\begin{matrix}
a^{zz} & = & z^{2}\cr a^{z\Lambda } & = & 0\cr a^{\Lambda \Sigma } & = &
z^{-1}\,\widetilde{G}^{\Lambda \Sigma }\cr
\end{matrix}
\right.  \label{aij1inv}
\end{equation}
Then, in this case the BH effective potential takes the form:
\begin{equation}
V^{(5)}=z^{2}e_{z}^{2}+z^{-1}\widetilde{G}^{\Lambda \Sigma }(X)e_{\Lambda
}e_{\Sigma }=z^{2}e_{z}^{2}+z^{-1}V^{(6)}(\hat{X})\,.  \label{potnovec}
\end{equation}
where $(e_{z},e_{\Lambda })\equiv q_{I}$ denote the electric charges and, to
obtain the last expression, we made use of (\ref{tildeg}). The physical
interpretation of the charges $e_{z}$ and $e_{\Lambda }$ is the following: $%
e_{z}$ is the Kaluza-Klein charge and $e_{\Lambda }$ are the charges of
dyonic strings wrapped around $S^{1}$.

The extrema of $V^{(5)}$ are found for:
\begin{equation}
\frac{\partial V^{(5)}}{\partial z}=0\Rightarrow 2ze_{z}^{2}-\frac{1}{2}%
z^{-2}V^{(6)}(\hat{X})=0
\end{equation}
which is the stabilization equation for the KK dilaton, solved by:
\begin{equation}
z=\left( \frac{V^{(6)}|_{extr}}{2e_{z}^{2}}\right) ^{\frac{1}{3}}
\end{equation}
and for:
\begin{equation}
\frac{\partial V^{(5)}}{\partial \hat{X}^{\Lambda }}=0\Rightarrow \frac{%
\partial V^{(6)}}{\partial \hat{X}^{\Lambda }}=0  \label{3.16}
\end{equation}
which shows that in this case the attractor solutions of the five
dimensional theory are precisely the same of the parent six dimensional
theory.

The BH entropy is now given by \cite{FG2}:
\begin{equation}
\left( S_{BH}^{(5)}\right) ^{4/3}=V^{(5)}|_{extr}=3\left( \frac{1}{2}%
e_{z}V^{(6)}|_{extr}\right) ^{\frac{2}{3}}=3\left( \frac{1}{2}%
e_{z}e^{\Lambda }e_{\Lambda }\right) ^{\frac{2}{3}}\,.
\end{equation}

The solution of Eqs. (\ref{3.16}) depends on whether the $d=6$ attractor is
BPS or not. As previously mentioned, the $d=6$ BPS attractors correspond to $%
Z_{i}=0$ $\forall i$, whereas the non-BPS ones are given by $Z=0$ (and $%
Z_{i}\neq 0$ for at least some $i$) \cite{ADFL-6,FG2}. Thus, all $q+1$ $d=6$
BPS moduli are fixed, while there are $q$ non-BPS flat directions, spanning
the $d=6$ non-BPS moduli space $\frac{SO\left( 1,q\right) }{SO\left(
q\right) }$ \cite{FG2}.

The supersymmetry-preserving features (BPS or non-BPS) of the $d=6$
attractors solutions depend on the sign of $e^{\Lambda }e_{\Lambda
}$: it is BPS for $e^{\Lambda }e_{\Lambda }>0$ and non-BPS for
$e^{\Lambda }e_{\Lambda }<0$. In this latter case, also the $d=5$
solution is non-BPS, because in a given frame \cite{AFL-6-5}
$e_{z}e^{\Lambda }e_{\Lambda }=e_{z}e_{+}e_{-}$ (with $e_{\pm
}\equiv e_{1}\pm e_{2}$), and if $e_{+}e_{-}<0$ the three
charges cannot have the same sign \cite{CFM1}. On the other hand, if $%
e^{\Lambda }e_{\Lambda }>0$ one can have both BPS and non-BPS $d=5$
solutions \cite{CFM1}.

Thus, we can conclude that for the ``generic sequence'' of $d=5$
symmetric real special spaces the non-BPS moduli space, predicted in
\cite {fm07}, does indeed coincide with the above mentioned $d=6$
(tensor multiplets') non-BPS moduli space, found in \cite{FG2}.



\setcounter{equation}0
\def\theequation{3.\arabic{subsection}.\arabic{equation}}

\subsection{\label{Subsect32}Inclusion of $d=6$ vector multiplets}

Let us now generalize the discussion to the case where $s$ extra vector
multiplets:
\begin{equation}
(A_{\mu },\lambda ^{A},Y)^{\alpha }\,,\qquad \alpha =1,\cdots ,s\,,
\end{equation}
corresponding to the dimensional reduction of six dimensional ones, are
present \cite{FMS}. The reduction may be done preserving the $SO(1,q+1)$
symmetry when the number $s$ of $d=6$ vector multiplets coincides with the
dimension of the spinor representation of $SO(1,q+1)$:
\begin{equation}
s=\dim \left[ \mbox{spin}\,SO(1,q+1)\right] .
\end{equation}
This implies that the kinetic matrix of the $d=6$ vector fields is positive
definite and no phase transitions, as discussed in \cite{Seiberg-Witten, FMS}%
, occur in this class of models.

The extra scalars contribute to the general relations (\ref{aij}) and (\ref
{veff5}) via a modification of the cubic form $\mathcal{V}$ into \cite
{vanderseypen}:
\begin{equation}
\mathcal{V}=\frac{1}{2}zX^{\Lambda }X^{\Sigma }\eta _{\Lambda \Sigma }+\frac{%
1}{2}X_{\Lambda }Y^{\alpha }Y^{\beta }\Gamma _{\alpha \beta }^{\Lambda }\,.
\label{cvvec}
\end{equation}
The total number of five dimensional scalars is then $q+2+s$. Of
particular interest are the four magic models which are associated
with the simple Jordan algebras having an irreducible norm form
(displayed in Table 4 of \cite{fm07}). In these cases $q=1,2,4,8$
and $s=2q$. Also the \textit{``}generic sequence'' $L(0,P)$ can be
viewed as a particular case of Eq. (\ref{cvvec}) with $q=0$ and
$s=P$.

The $d=6$ origin of the second term in Eq. (\ref{cvvec}) is the
kinetic term of the $d=6$ vector fields, which reads \cite{Sagnotti,
FMS} ($\Lambda =0,1,...,q+1$, $\alpha =1,...,s$, $C_{\alpha \beta
}^{\Lambda }=C_{\beta \alpha }^{\Lambda }$)
\begin{equation}
X_{\Lambda }C_{\alpha \beta }^{\Lambda }F^{\alpha }\wedge {}^*
F^{\beta }\,.
\end{equation}
Thus, in the presence of $d=6$ BH charges $Q^{\alpha }$, it originates an
effective $d=6$ BH potential of the form
\begin{equation}
V_{BH}^{(6)}=X_{\Lambda }C_{\alpha \beta }^{\Lambda }Q^{\alpha }Q^{\beta }.%
\label{VBH-6}
\end{equation}
Such a potential has run-away extrema at $d=6$ \cite{adf96}. This
can be seen for instance in the case $n_{T}=1\Leftrightarrow q=0$,
where Eq. (\ref{VBH-6}) reduces to ($\alpha =1,...,P$,
$X_{0}=cosh\phi $, $X_{1}=sinh\phi $)
\begin{equation}
V_{BH}^{(6)}\left( \phi \right) =cosh\phi \,C_{\alpha \beta
}^{0}Q^{\alpha }Q^{\beta }+sinh\phi \, C_{\alpha \beta
}^{1}Q^{\alpha }Q^{\beta }=e^{\phi }Q^{\alpha }Q^{\alpha },
\end{equation}
 (in the last step we used the fact that in the $n_T=1$ case we may set $C_{\alpha \beta
}^{0}=C_{\alpha \beta }^{1}=\delta _{\alpha \beta }$ without loss of
generality). Consequently
\begin{equation}
\frac{\partial V_{BH}^{(6)}\left( \phi \right) }{\partial \phi }%
=0\Leftrightarrow V_{BH}^{(6)}\left( \phi \right) =0\Leftrightarrow \phi
=-\infty .
\end{equation}
We then conclude that, besides BPS  BH attractors, also non-BPS
extremal BH attractors are  excluded in $(1,0)$ supergravity in six
dimensions. However,  we can have a $0$-dimensional black object by
an intersection  of a $d=6$ BH with a $d=6$ black string. Its
reduction to $d=5$ gives a BH which carries  both the string charge
and the BH charge, with cubic invariant of the form \cite{FM}
\begin{equation}
I_{3}=e_{z}e^{\Lambda }e_{\Lambda }+e_{\Lambda }C_{\alpha \beta }^{\Lambda
}Q^{\alpha }Q^{\beta },\label{I3-I3}
\end{equation}
and  $d=5$ resulting BH entropy  $S_{BH}^{(5)}\sim \sqrt{\left|
I_{3}\right| }$. Thus, even if the KK charge $e_{z}$ vanishes, one
gets a contribution from the second term of Eq. (\ref{I3-I3}). This
is in contrast with the case of the $d=6$ dyonic extremal black
string treated in Subsect. \ref{Subsect31}, where the non-vanishing
of the KK charge $e_{z}$ was needed in order to get a non-vanishing
entropy for the corresponding $d=5 $ BH, obtained by wrapping the
$d=6$ string on $S^{1}$.

The inclusion of extra multiplets corresponding to $d=6$ vector
multiplets induces a significative complication in the model. In
particular, the moduli space of the non-BPS attractors drastically
changes with respect to the case described in section \ref
{Subsect31}. As we shall prove below, in the magic models the number
of moduli becomes equal to $s=2q$ instead of $q$ as it was in the
absence of these extra multiplets.

Before entering into the detail of the magic models, let us argue the
existence, at least for the homogeneous spaces $L(q,P)$ (and, for $q=4m$, $%
L(q,P,P^{\prime })$) \cite{vanderseypen}, of particular non-BPS critical
points where the same results of section \ref{Subsect31} may still be
directly applied. Indeed, it turns out that for the four magic models the
non-BPS attractor moduli spaces of dimension $2q$ always contain as a
subspace precisely the coset $\frac{SO(1,q)}{SO(q)}$ (that is the moduli
space of $d=6$ non-BPS attractors for $q+1$ strings, as discussed above).
Such submanifold of the moduli space may be obtained by considering the
particular critical point where $Y^{\alpha }=0$. This critical point may
always be reached because, as (\ref{cvvec}) and (\ref{aij}) show, the $Y$
coordinates always appear quadratically in the effective potential (\ref
{veff5}). Then, for $Y^{\alpha }=0$ the effective potential reduces to the
one previously considered (Eq. (\ref{potnovec})), whose non-BPS attractor
solution is known to have $q$ flat directions belonging to the coset $\frac{%
SO(1,q)}{SO(q)}$. This is in fact only half the total number of flat
directions for these solutions. It may be understood because the non
compact stabilizer of the non-BPS orbit (that is for example
$F_{4(-20)}\supset SO(1,8)$ for $q=8$ \cite{helgason,gilmore}),
mixes the $X$ with $Y$ variables, so that the restriction
$\{Y^{\alpha }\}=0$ implies the reduction of the orbit to its
subgroup $SO(1,q)$. The same considerations may be directly
extended, for charge configurations where the spinorial charges are
set to zero, to the series of homogeneous non-symmetric spaces
$L(q,P)$ (and, for $q=4m$, $L(q,P,P^{\prime })$)
\cite{vanderseypen}, which always admit a non-BPS attractor point
where all the spinorial moduli are zero. As before, this condition
selects the submanifold $\frac{SO(1,q)}{SO(q)}$ of the non-BPS
attractor moduli space, with the only difference that in this case
the number $q$ is not directly related to the number of spinorial
moduli.


\setcounter{equation}0
\def\theequation{3.2.\arabic{subsubsection}.\arabic{equation}}

\subsubsection{\label{Subsubsect321}$N=2$ magic models}

For $N=2$ supergravity, one can apply the general relations of real special
geometry \cite{gst,FG2}, so that the efffective potential
\begin{equation}
V(\phi ,q)=a^{IJ}q_{I}q_{J}
\end{equation}
takes a simpler form. Indeed, for $N=2$ supergravity the vector kinetic
matrix ${a}_{IJ}$ is related to the metric $g_{xy}$ of the scalar manifold
via
\begin{equation}
{a}_{IJ}=h_{I}h_{J}+\frac{3}{2}h_{I,x}h_{J,y}g^{xy}  \label{eq:vectormetric}
\end{equation}
$a^{IJ}=h^{I}h^{J}+\frac{3}{2}h_{,x}^{I}h_{,y}^{J}g^{xy}$ or conversely
\begin{equation}
g_{xy}=\frac{3}{2}h_{I,x}h_{J,y}{a}^{IJ}\,.
\end{equation}
In terms of these quantities the central charge is
\begin{equation}
Z=q_{I}h^{I}
\end{equation}
and we can write the potential as
\begin{equation}
V(q,\phi )=Z^{2}+\frac{3}{2}g^{xy}\partial _{x}Z\partial _{y}Z  \label{eq:N2}
\end{equation}
where $\partial _{x}Z=q_{I}h_{,x}^{I}=P_{x}^{a}Z_{a}$ are the matter
charges. The index $x=1,\cdots ,n_{V}$ is a world index labelling the scalar
fields while $a$ is the corresponding rigid index. $P_{x}^{a}$ denotes the
scalar vielbein. The matter charges obey the differential relations:
\begin{eqnarray}
\nabla Z &=&P^{a}\,Z_{a}\,;  \notag \\
\nabla Z_{a} &=&\frac{2}{3}\,g_{ab}\,P^{b}\,Z-\sqrt{\frac{2}{3}}%
T_{abc}\,P^{b}\,g^{cd}Z_{d}.  \label{nabZ}
\end{eqnarray}
To make explicit computations of the attractor points of the potential and
of the corresponding Hessian matrix, let us use the property that both $%
T_{abc}$ and $g_{ab}$, written in rigid indices, are invariants of the group
$\mathrm{SO}(q+1)$, where $q=1,2,4,8$ for the magic models, corresponding to
the symmetric spaces $L(q,1)$. The $H$--representation $\mathbf{R%
}$ of the scalar fields branch with respect to $\mathrm{SO}(q+1)$ in the
following way
\begin{equation}
\mathbf{R}\rightarrow \mathbf{1}+(\mathbf{q+1})+\mathbf{R_{s}}\,,
\end{equation}
where $\mathbf{R_{s}}$ is the real Clifford module of $\mathrm{SO}(q+1)$ of
dimensions $\mathrm{dim}(\mathbf{R_{s}})=2,4,8,16$ corresponding to the four
values of $q$. The index $a$ split into the indices $1,\,m,\,\alpha $, where
$m=1,\dots ,q+1$ and $\alpha =1,\dots ,\mathrm{dim}(\mathbf{R_{s}})$. Let us
write the general form for $T_{abc}$ and $g_{ab}$:
\begin{eqnarray}
g_{11} &=&\alpha \,\,;\,\,\,g_{mn}=\beta \,\delta _{mn}\,\,;\,\,\,g_{\alpha
\beta }=\gamma \,\delta _{\alpha \beta }\,;  \notag \\
T_{111} &=&\sqrt{\frac{\alpha}{2}}\,g_{11}\,\,;\,\,\,T_{1mn}=-
 \,\sqrt{\frac{\alpha }{2}}\,\,g _{mn}\,\,;\,\,\,T_{1\alpha \beta
}=\frac{1}{2}\, \,\sqrt{\frac{\alpha }{2}}\, g_{\alpha \beta }\,;
\notag \\
T_{n\alpha \beta } &=&-\frac{1}{2}\,\gamma \,\sqrt{\frac{3}{2}\,\beta }%
\,\,\Gamma _{n\alpha \beta }\,,
\end{eqnarray}
where $\Gamma _{n}$ are the (symmetric, real) $\mathrm{SO}(q+1)$ gamma
matrices in the $\mathbf{R_{s}}$ representation. The coefficients of $T_{abc}
$ are determined in terms of the coefficients of $g_{ab}$ by the following
relation:
\begin{equation}
T_{a(bc}T^{a}{}_{ef)}=\frac{1}{2}\,g_{(bc}\,g_{ef)}\,.
\end{equation}
The potential $V$ can be written in the following useful form:
\begin{equation}
V=Z^{2}+\frac{3}{2}g^{ab}\,Z_{a}\,Z_{b}=Z^{2}+\frac{3}{2}\left(
Z_{1}Z^{1}+Z_{n}Z^{n}+Z_{\alpha }Z^{\alpha }\right) \,,
\end{equation}
where the following short-hand notation is used: $Z^{a}\equiv g^{ab}Z_{b}$.
Let us now compute the extrema of $V$. Using eqs. (\ref{nabZ}) we find
\begin{eqnarray}
\nabla V &=&P^{1}\,\left[ 4\,Z\,Z_{1}-\sqrt{3\alpha }\left(
Z_{1}Z^{1}-Z_{n}Z^{n}+\frac{1}{2}Z_{\alpha }Z^{\alpha }\right) \right] +\,
\notag \\
&&+P^{n}\,\left( 4\,Z\,Z_{n}+2\,\sqrt{\frac{3}{\alpha }}\,Z_{1}\,Z_{n}+\frac{%
3}{2\,\gamma }\,\sqrt{\beta }\,\Gamma _{n\alpha \beta }\,Z_{\alpha
}\,Z_{\beta }\right) +  \notag \\
&&+P^{\alpha }\,\left( 4\,Z\,Z_{\alpha }-\frac{3}{2}\,\sqrt{\frac{3}{\alpha }%
}\,Z_{1}\,Z_{\alpha }+\frac{3}{\sqrt{\beta }}\,\Gamma _{n\alpha \beta
}\,Z_{n}\,Z_{\beta }\right) \,.
\end{eqnarray}
It is straightforward to see that the above expression has two zeroes
corresponding to the two attractors:

\begin{itemize}
\item  {\ \textbf{BPS attractor:} $Z_n=Z_\alpha=Z_1=0$ and the potential at
the extremum reads $V_0=Z^2$;}

\item  {\textbf{non-BPS attractor:} $Z_n=Z_\alpha=0$, $Z=\frac{1}{4}\,\sqrt{%
\frac{3}{\alpha}}\,Z_1$ and the potential at the extremum reads $V_0=9\,Z^2 $%
.}
\end{itemize}

Let us now compute the Hessian matrix:
\begin{eqnarray}
\nabla ^{2}V &=&(P^{1})^{2}\,\left\{ \frac{8\alpha }{3}\left[ \left( Z-2\,%
\frac{1}{4}\sqrt{\frac{3}{\alpha }}Z_{1}\right) ^{2}+8\left( \frac{1}{4}%
\sqrt{\frac{3}{\alpha }}Z_{1}\right) ^{2}\right] +2\,\frac{\alpha }{\beta }%
\,Z_{n}^{2}+\frac{\alpha }{2\gamma }Z_{\alpha }^{2}\right\} +  \notag \\
&&+P^{1}P^{n}\,\left( 8\,Z_{1}\,Z_{n}+16\,\sqrt{\frac{\alpha }{3}}\,Z\,Z_{n}-%
\frac{\sqrt{3\alpha \beta }}{\gamma }\,\Gamma _{n\alpha \beta }\,Z_{\alpha
}\,Z_{\beta }\right) +  \notag \\
&&+P^{1}P^{\alpha }\,\left( 11\,Z_{1}\,Z_{\alpha }-8\,\sqrt{\frac{\alpha }{3}%
}\,Z\,Z_{\alpha }+\sqrt{\frac{3\alpha }{\beta }}\,\Gamma _{n\alpha \beta
}\,Z_{n}\,Z_{\beta }\right) +  \notag \\
&&+P^{n}P^{m}\,\left\{ 6\,Z_{n}\,Z_{m}+\left[ \frac{8\beta }{3}\left( Z+2\,%
\frac{1}{4}\sqrt{\frac{3}{\alpha }}Z_{1}\right) ^{2}+\frac{3\beta }{2\gamma }%
Z_{\alpha }^{2}\right] \,\delta _{mn}\right\} +  \notag \\
&&+P^{n}P^{\alpha }\,\left( 6\,Z_{n}\,Z_{\alpha }+8\,\sqrt{\beta }\,\Gamma
_{n\alpha \beta }\,ZZ_{\alpha }+\sqrt{\frac{3\beta }{\alpha }}\,\Gamma
_{n\alpha \beta }\,Z_{\beta }Z_{1}+3\,(\Gamma _{m}\,\Gamma _{n})_{\alpha
\beta }Z_{m}\,Z_{\beta }\right) +  \notag \\
&&+P^{\alpha }P^{\beta }\,\left[ \frac{8}{3}\,\gamma \,\left( Z-\frac{1}{4}\,%
\sqrt{\frac{3}{\alpha }}\,Z_{1}\right) ^{2}\,\delta _{\alpha \beta }+\frac{%
3\gamma }{2\beta }\,Z_{n}^{2}\,\delta _{\alpha \beta }+\frac{4\gamma }{\sqrt{%
\beta }}\,\Gamma _{n\alpha \beta }\,ZZ_{n}-\right.  \notag \\
&&-\left. \gamma \,\sqrt{\frac{3}{\alpha \beta }}\,\Gamma _{n\alpha \beta
}Z_{1}Z_{n}+\frac{3}{2}\,\Gamma _{n\alpha \delta }\,\Gamma _{n\beta \gamma
}\,Z_{\delta }Z_{\gamma }+\frac{9}{2}\,Z_{\alpha }Z_{\beta }\right]
\end{eqnarray}
At the BPS critical point it is straightforward to check that:
\begin{equation}
\nabla ^{2}V=\frac{8}{3}Z^{2}\,g_{ab}\,P^{a}\,P^{b}\,.
\end{equation}
As expected, the BPS critical point is a stable attractor. At the non-BPS
attractor the Hessian reads:
\begin{equation}
\nabla ^{2}V=24\,Z^{2}\,\left[ \,g_{11}(P^{1})^{2}+g_{mn}\,P^{n}P^{m}\right]
\,.
\end{equation}
The moduli space is therefore spanned by the scalar fields in the $\mathbf{%
R_{s}}$ representation. These can be regarded as particular coordinates of
the moduli spaces of the $N=2$, $d=5$ non-BPS solutions of the magic models $%
J_{3}^{\mathbb{O}}$, $J_{3}^{\mathbb{H}}$, $J_{3}^{\mathbb{C}}$ and $J_{3}^{%
\mathbb{R}}$, which respectively are $\frac{F_{4(-20)}}{SO(9)}$, $\frac{%
USp(4,2)}{USp(4)\times USp(2)}$, $\frac{SU(2,1)}{SU(2)\times U(1)}$ and $%
\frac{SL\left( 2,\mathbb{R}\right) }{SO(2)}$ (see Table 4 of \cite{fm07}).
It is worth pointing out that, with the exception of $J_{3}^{\mathbb{O}}$,
all such spaces can be obtained as consistent truncations of the $N=8$, $d=5$
BPS attractor moduli space $\frac{F_{4(4)}}{USp(6)\times USp(2)}$
(quaternionic K\"{a}hler), by performing an analysis which is the $d=5$
counterpart of the $d=4$ analysis exploited in \cite{Ferrara-Marrani-1}.
Since for $J_{3}^{\mathbb{C}}$ and $J_{3}^{\mathbb{R}}$ the $%
N=8\longrightarrow N=2$ reduction preserves $n_{H}=1$ and $n_{H}=2$
hypermultiplets respectively, the following inclusions must hold:
\begin{equation}
J_{3}^{\mathbb{C}}:F_{4(4)}\supset \left( SU\left( 2,1\right) \right)
^{2}\Longrightarrow \frac{F_{4(4)}}{USp(6)\times USp(2)}\supset \frac{SU(2,1)%
}{SU(2)\times U(1)}\times \frac{SU(2,1)}{SU(2)\times U(1)};  \label{J3C-incl}
\end{equation}
\begin{equation}
J_{3}^{\mathbb{R}}:F_{4(4)}\supset SL\left( 2,\mathbb{R}\right) \times
G_{2(2)}\Longrightarrow \frac{F_{4(4)}}{USp(6)\times USp(2)}\supset \frac{%
SL(2,\mathbb{R})}{SO(2)}\times \frac{G_{2(2)}}{SO(4)}.  \label{J3R-incl}
\end{equation}
The two group embeddings given by Eqs. (\ref{J3C-incl}) and (\ref{J3R-incl})
are discussed in Appendix.

On the other hand, the truncation generating $J_{3}^{\mathbb{H}}$ implies
\begin{equation}
J_{3}^{\mathbb{H}}:F_{4(4)}\supset USp\left( 4,2\right) \Longrightarrow
\frac{F_{4(4)}}{USp(6)\times USp(2)}\supset \frac{USp(4,2)}{USp(4)\times
USp(2)}.  \label{J3H-incl}
\end{equation}
In this case, the $\mathbf{42}$ of $USp\left( 8\right) $ decomposes along $%
USp\left( 6\right) \times USp\left( 2\right) $ as $\mathbf{42\longrightarrow
}\left( \mathbf{14},\mathbf{1}\right) \oplus \left( \mathbf{14}^{\prime },%
\mathbf{2}\right) $. The $\mathbf{14}$ and $\mathbf{14}^{\prime }$ of $%
USp\left( 6\right) $ further decompose with respect to $USp\left( 4\right)
\times USp\left( 2\right) $ (maximal compact subgroup of the stabilizer $%
USp\left( 4,2\right) $ of the non-BPS orbit) as follows:
\begin{equation}
\begin{array}{l}
\mathbf{14\longrightarrow }\left( \mathbf{1},\mathbf{1}\right) \oplus \left(
\mathbf{5},\mathbf{1}\right) \oplus \left( \mathbf{4},\mathbf{2}\right) ; \\
\\
\mathbf{14}^{\prime }\mathbf{\longrightarrow }\left( \mathbf{5},\mathbf{2}%
\right) \oplus \left( \mathbf{4},\mathbf{1}\right) .
\end{array}
\end{equation}
Thus, the decomposition of the $\left( \mathbf{14},\mathbf{1}\right) $ and $%
\left( \mathbf{14}^{\prime },\mathbf{2}\right) $ of $USp\left( 6\right)
\times USp\left( 2\right) $ with respect to $USp\left( 4\right) \times
USp\left( 2\right) \times USp\left( 2\right) $ read:
\begin{equation}
\begin{array}{l}
massive:\left( \mathbf{14},\mathbf{1}\right) \mathbf{\longrightarrow }\left(
\mathbf{1},\mathbf{1},\mathbf{1}\right) \oplus \left( \mathbf{5},\mathbf{1},%
\mathbf{1}\right) \oplus \left( \mathbf{4},\mathbf{2},\mathbf{1}\right) ; \\
\\
massless:\left( \mathbf{14}^{\prime },\mathbf{2}\right) \mathbf{%
\longrightarrow }\left( \mathbf{5},\mathbf{2},\mathbf{2}\right) \oplus
\left( \mathbf{4},\mathbf{1},\mathbf{2}\right) .
\end{array}
\label{CERN-17-09}
\end{equation}
Since in the non-BPS case the $N=2$ $\mathcal{R}$-symmetry is the $USp\left(
2\right) \sim SU(2)$ inside $USp\left( 6\right) $ (\textit{i.e.} the first $%
USp\left( 2\right) $ in the decomposition (\ref{CERN-17-09})) one obtains $8$
massive and $20$ massless hypermultiplets' degrees of freedom, and $6$
massive and $8$ massless vectors' degrees of freedom. Notice that, since in
the BPS case the $N=2$ $\mathcal{R}$-symmetry is the $USp\left( 2\right)
\sim SU(2)$ commuting with $USp\left( 6\right) $ (\textit{i.e.} the second $%
USp\left( 2\right) $ in the decomposition (\ref{CERN-17-09})), the non-BPS
case differs from the BPS case only by an exchange of the $\left( \mathbf{4},%
\mathbf{2},\mathbf{1}\right) $ representation with the $\left( \mathbf{4},%
\mathbf{1},\mathbf{2}\right) $ one. \setcounter{equation}0
\def\theequation{\arabic{section}.\arabic{equation}}

\section{\label{Sect4}Purely five dimensional analysis of attractors in $N$%
-extended theories}

For any extended supergravity in five dimensions the BH potential
enjoys the general expression in terms of the dressed charges \cite
{adf5d,ADF-U-duality}:
\begin{equation}
V(\phi ,q)=\frac{1}{2}Z_{AB}Z^{AB}+X^{2}+Z_{I}Z^{I}
\end{equation}
where $Z_{AB}$ ($A,B=1,\cdots N$) are the antisymmetric, $Sp(N)$-traceless
graviphoton central charges, $X$ the trace part while $Z_{I}$ ($I=1,\cdots
,n $) denote the matter charges (which only appear for $N\leq 4$ theories).
For all the models with a scalar sector spanning a symmetric space, the
dressed charges obey some known differential relations in moduli space which
allow to explicitly find the attractor condition as an extremum for the
scalar potential in moduli space:
\begin{equation}
\frac{\partial V}{\partial \phi ^{i}}=0\,.
\end{equation}

We are going to study in the following the BPS and non-BPS attractors for
the various cases.

\setcounter{equation}0
\def\theequation{4.\arabic{subsection}.\arabic{equation}}

\subsection{\label{Subsect41}$N=8$, $d=5$ and $(2,2)$, $d=6$}

The scalar manifold is the coset
\begin{equation}
G/H=\frac{E_{6(6)}}{Sp(8)}\,.  \label{n=8coset}
\end{equation}
and the BH potential takes the form:
\begin{equation}
V=\frac{1}{2}Z_{AB}Z^{AB}\,.  \label{V-N=8}
\end{equation}
The differential relations among the 27 central charges $Z_{AB}$ (satisfying
$Z_{AB}\,\Omega ^{AB}=0$), are:
\begin{equation}
\nabla Z_{AB}=\frac{1}{2}Z^{CD}P_{ABCD}\,,  \label{diffrel8}
\end{equation}
where the vielbein $P_{ABCD}=P_{ABCD,i}d\phi ^{i}$ satisfies the conditions
\begin{equation}
P^{ABCD}=P^{[ABCD]}\,,\qquad P^{ABCD}\Omega _{AB}=0\,.
\end{equation}
The extremum condition is then
\begin{equation}
\nabla V=Z_{AB}\nabla Z^{AB}=\frac{1}{2}P^{ABCD}Z_{AB}Z_{CD}\,=0.
\label{extrn=8}
\end{equation}
To explicitly find the solution, it is convenient to put the central-charge
matrix in normal form:
\begin{equation}
Z_{AB}=
\begin{pmatrix}
e_{1} & 0 & 0 & 0\cr0 & e_{2} & 0 & 0\cr0 & 0 & e_{3} & 0\cr0 & 0 & 0 &
-e_{1}-e_{2}-e_{3}
\end{pmatrix}
\otimes
\begin{pmatrix}
0 & 1\cr-1 & 0
\end{pmatrix}
\,,  \label{z8}
\end{equation}
and to truncate the theory to the ``charged'' submanifold spanned by the
vielbein components that couple to the dressed charge in normal form, that
is:
\begin{eqnarray}
&&P_{1}\equiv P_{1234}=P_{5678}\,,\qquad P_{2}\equiv P_{1256}=P_{3478}
\notag \\
&&\mbox{while }P_{3456}=P_{1278}=-P_{1}-P_{2}\,.
\end{eqnarray}
In this way, the covariant derivatives of the charges (\ref{diffrel8})
become:
\begin{eqnarray}
\nabla e_{1} &=&(e_{1}+2e_{2}+e_{3})P_{1}+(e_{1}+e_{2}+2e_{3})P_{2}  \notag
\label{difnor8} \\
\nabla e_{2} &=&(e_{1}-e_{3})P_{1}+(-e_{1}-e_{2}-2e_{3})P_{2}  \notag \\
\nabla e_{3} &=&(-e_{1}-2e_{2}-e_{3})P_{1}+(e_{1}-e_{2})P_{2}\,.
\end{eqnarray}
Using these relations, the extremum condition of $V$ becomes
\begin{equation}
\nabla V=4\left\{
P_{1}(e_{1}-e_{3})(e_{1}+2e_{2}+e_{3})+P_{2}(e_{1}-e_{2})(e_{1}+e_{2}+2e_{3})\right\} =0.
\end{equation}
It admits only one solution with finite area, which breaks the symmetry $%
Sp(8)\rightarrow Sp(2)\times Sp(6)$. Up to $Sp(6)$ rotations it is:
\begin{equation}
e_{2}=e_{3}=-{\frac{1}{3}}e_{1}\,;\quad V_{extr}={\frac{4}{3}}e_{1}^{2}={%
\frac{4}{3}}M_{extr}^{2}\,.  \label{extrz8}
\end{equation}
This is a BPS attractor, supported by the unique BPS orbit \cite{FG2} $\frac{%
E_{6(6)}}{F_{4(4)}}$, and the maximum amount of supersymmetry preserved by
the solution at the horizon is 1/4 ($N=8\rightarrow N=2$).

As mentioned above, the two vielbein-components $P_{1}$ and $P_{2}$ span the
submanifold of the moduli space which couples to the proper values of the
central charge. This automatically projects out, in the $N=2$ reduced
theory, the 28 scalar degrees of freedom corresponding to the
hypermultiplets.

The Hessian matrix reads
\begin{equation}
H_{ij}\equiv \nabla _{i}\nabla
_{j}V=\frac{1}{4}P_{ABLM}P^{CDLM}Z^{AB}Z_{CD}.  \label{h8}
\end{equation}

To have the complete spectrum of massive plus flat directions, we have to
consider in (\ref{h8}) the complete vielbein $P_{ABCD}$. On the solution,
where $Sp(8)\rightarrow Sp(2)\times Sp(6)$ ($A\rightarrow (\alpha ,a)$, $%
\alpha =1,2$, $a=1,\cdots ,6$), the vielbein degrees of freedom decompose as
\begin{eqnarray}
\mathbf{42} &\rightarrow &\mathbf{(14,1)}+\mathbf{(14^{\prime },2)}  \notag
\\
P_{ABCD} &\rightarrow &P_{\alpha \beta ab}+P_{\alpha abc}
\end{eqnarray}
where $P_{\alpha \beta ab}=\epsilon _{\alpha \beta }P_{ab}$ (satisfying $%
P_{ab}\Omega ^{ab}=0$) is the vielbein of the $\frac{SU^{\ast }\left(
6\right) }{Sp\left( 6\right) }$ $N=2$ vector multiplet sigma model, while $%
P_{\alpha abc}$ (satisfying $P_{\alpha abc}\Omega ^{ab}=0$) spans the $N=2$
hyperscalar sector. Note that, at the horizon, from (\ref{z8}) and (\ref
{extrz8}) we find, for the central charge in normal form:
\begin{equation}
Z_{AB}\rightarrow \quad (Z_{ab}=e\,\Omega _{ab}\,;\quad Z_{\alpha \beta
}=-3\,e\,\epsilon _{\alpha \beta })
\end{equation}
The Hessian matrix (\ref{h8}) is then:
\begin{equation}
H_{ij}=\frac{1}{4}\left( P_{abLM}Z^{ab}+P_{\alpha \beta LM}Z^{\alpha \beta
}\right) \left( P^{cdLM}Z_{cd}+P^{\gamma \delta LM}Z_{\gamma \delta }\right)
=9e^{2}P_{LM,i}P_{,j}^{LM}\,.  \label{h8big}
\end{equation}
The hyperscalar vielbein $P_{\alpha abc}$ do not appear in
(\ref{h8big}) so that the corresponding directions do not acquire a
mass. The moduli space of the solution is then \cite{fm07}
$\frac{F_{4(4)}}{USp(6)\otimes USp(2)}$.

The $N=8$, $d=5$ theory has an uplift to $\left( 2,2\right) $, $d=6$
supergravity, whose scalar manifold is $\frac{SO\left( 5,5\right)
}{SO\left( 5\right) \times SO\left( 5\right) }$ \cite{ADFL-6}. In
such a theory, the
unique orbit with non-vanishing area is the $\frac{1}{4}$-BPS orbit $\frac{%
SO\left( 5,5\right) }{SO\left( 5,4\right) }$ \cite{LPS}, specified by an $%
SO\left( 5,5\right) $ charge vector $e_{\Lambda }$ with non-vanishing norm $%
e_{\Lambda }e^{\Lambda }\neq 0$. The corresponding moduli space of $\frac{1}{%
4}$-BPS attractors is $\frac{SO\left( 5,4\right) }{SO\left( 5\right)
\times
SO\left( 4\right) }$, and it is indeed contained \cite{gilmore} in the $N=8$%
, $d=5$ $\frac{1}{8}$-BPS moduli space $\frac{F_{4\left( 4\right) }}{%
USp\left( 6\right) \times USp\left( 2\right) }$, as implied by our
analysis.
Note that the two non-compact forms of $F_{4}$ which occur in $N=2$ and $N=8$%
, $d=5$ supergravities precisely contain the two non-compact forms of $%
SO\left( 9\right) $ present in the corresponding moduli spaces \cite{gilmore}%
: $F_{4(-20)}\supset SO\left( 1,8\right) $ and $F_{4(4)}\supset
SO\left( 5,4\right) $.\setcounter{equation}0
\def\theequation{4.\arabic{subsection}.\arabic{equation}}

\subsection{\label{Subsect42}$N=6$ ($N=2,~J_{3}^{\mathbb{H}}$)}

The scalar manifold is the coset
\begin{equation}
G/H=\frac{SU^{\ast }(6)}{USp(6)}\,,  \label{n=6coset}
\end{equation}
the BH potential takes the form:
\begin{equation}
V=\frac{1}{2}Z_{AB}Z^{AB}+\frac{1}{3}X^{2}\,,
\end{equation}
and the differential relations among the 14+1 central charges $Z_{AB}$
(satisfying $Z_{AB}\,\Omega ^{AB}=0$) and $X$, are:
\begin{eqnarray}
\nabla Z_{AB} &=&\Omega ^{CD}Z_{C[A}P_{B]D}+\frac{1}{6}\Omega
_{AB}Z_{CD}P^{CD}+\frac{1}{3}XP_{AB}  \notag \\
\nabla X &=&\frac{1}{2}Z_{AB}P^{AB}\,,  \label{diffrel6}
\end{eqnarray}
where $P_{AB}=P_{AB,i}d\phi ^{i}$ is the $\Omega $-traceless vielbein of $%
G/H $ satisfying the conditions
\begin{equation}
P^{AB}=P^{[AB]}\,,\qquad P^{AB}\Omega _{AB}=0\,.
\end{equation}
To study the attractors, it is convenient to put the central-charge matrix
in normal form:
\begin{equation}
Z_{AB}=
\begin{pmatrix}
e_{1} & 0 & 0\cr0 & e_{2} & 0\cr0 & 0 & -e_{1}-e_{2}
\end{pmatrix}
\otimes
\begin{pmatrix}
0 & 1\cr-1 & 0
\end{pmatrix}
\,,  \label{z6}
\end{equation}
so that the BH potential takes the form
\begin{equation}
V=e_{1}^{2}+e_{2}^{2}+(e_{1}+e_{2})^{2}+\frac{1}{3}X^{2}\,.
\end{equation}
The vielbein components that couple to the dressed charges in normal form
are:
\begin{eqnarray}
&&P_{1}\equiv P_{12}\,,\qquad P_{2}\equiv P_{34}  \notag \\
&&\mbox{while }P_{56}=-P_{1}-P_{2}\,.
\end{eqnarray}
In this way, the covariant derivatives of the charges (\ref{diffrel6})
become:
\begin{eqnarray}
\nabla e_{1} &=&\frac{1}{3}(-e_{1}+e_{2}+X)P_{1}+\frac{1}{3}%
(e_{1}+2e_{2})P_{2}  \notag  \label{difnor6} \\
\nabla e_{2} &=&\frac{1}{3}(2e_{1}+e_{2})P_{1}+\frac{1}{3}%
(e_{1}-e_{2}+X)P_{2}  \notag \\
\nabla X &=&(2e_{1}+e_{2})P_{1}+(e_{1}+2e_{2})P_{2}\,.
\end{eqnarray}

Using these relations, the extremum condition of $V$ becomes
\begin{eqnarray}
\nabla V&=&\frac 23 X Z_{AB}P^{AB} + \Omega^{CD} Z_{CA}Z^{AB} P_{BD}  \notag
\\
&=&2 \Bigl\{P_1 (2e_1+e_2)(e_2 +\frac 23 X) +P_2 (e_1+2e_2)(e_1+\frac 23
X)\Bigr\}=0
\end{eqnarray}
Two inequivalent solutions with finite area are there:

\begin{enumerate}
\item  $e_{1}=e_{2}=-\frac{2}{3}X$, giving for the Bekenstein--Hawking
entropy $V_{extr}=3X^{2}$.\newline
This is the $N=6$ 1/6-BPS solution and breaks the symmetry of the theory to $%
Sp(4)\times Sp(2)$.

\item  $e_{1}=e_{2}=0$, with Bekenstein--Hawking entropy $V_{extr}=\frac{1}{3%
}X^{2}$.\newline
It is a non-BPS attractor of the $N=6$ theory, and leaves all the $Sp(6)$
symmetry of the theory unbroken.
\end{enumerate}

Since the bosonic sector of this theory coincides with the one of an $N=2$
theory based on the same coset space \cite{FG2}, these are also the
attractor solutions of the corresponding $N=2$ model. In the $N=2$ version,
however, the interpretation of the attractor solutions as BPS and non-BPS
are interchanged.

To study the stability of the solutions, let us consider the Hessian matrix
\begin{eqnarray}
H_{ij} &\equiv &\nabla _{i}\nabla _{j}V  \notag  \label{h6} \\
&=&P_{AB}P_{CD}\Bigl[Z^{AC}Z^{BD}+\frac{2}{9}X^{2}\Omega ^{AC}\Omega ^{BD}-%
\frac{4}{3}XZ^{AC}\Omega ^{BD}+  \notag \\
&&+Z^{AL}Z_{LM}\Omega ^{BC}\Omega ^{MD}\Bigr]  \notag \\
&=&P_{AB}P_{CD}\left( Z^{AC}-\frac{1}{3}X\Omega ^{AC}\right) \left( Z^{BD}-%
\frac{1}{3}X\Omega ^{BD}\right) +  \notag \\
&&-P_{AB}P^{DB}\left( Z^{AC}-\frac{1}{3}X\Omega ^{AC}\right) \left( Z_{CD}-%
\frac{1}{3}X\Omega _{CD}\right)
\end{eqnarray}
and evaluate it on the two extrema. In the first case (BPS $N=6$, non-BPS $%
N=2$) the solution breaks the symmetry to $Sp(4)\times Sp(2)$, ($%
A\rightarrow (\alpha ,a)$, $\alpha =1,2$, $a=1,\cdots ,4$), since at the
horizon we find, for the central charge in normal form:
\begin{equation}
Z_{AB}\rightarrow \quad (Z_{ab}=-\frac{2}{3}\,X\,\Omega _{ab}\,;\quad
Z_{\alpha \beta }=\frac{4}{3}\,X\,\epsilon _{\alpha \beta })\,,
\end{equation}
so that
\begin{equation}
Z_{AB}-\frac{1}{3}X\Omega _{AB}\rightarrow \left\{
\begin{matrix}
\quad Z_{ab}-\frac{1}{3}X\Omega _{ab}=-X\,\Omega _{ab}\cr Z_{\alpha \beta }-%
\frac{1}{3}X\epsilon _{\alpha \beta }=X\,\epsilon _{\alpha \beta }\,.
\end{matrix}
\right.
\end{equation}
Corresponding to the group decomposition of the degrees of freedom:
\begin{eqnarray}
\mathbf{14} &\rightarrow &\mathbf{(5,1)}+\mathbf{(1,1)}+\mathbf{(4,2)}
\notag \\
P_{AB} &\rightarrow &(P_{ab};\,P;\,P_{a\alpha })\,,
\end{eqnarray}
the scalar vielbein decomposes as
\begin{equation}
P_{AB}\rightarrow \left\{
\begin{matrix}
\epsilon _{\alpha \beta }P\hfill \cr P_{\alpha a}\equiv -P_{a\alpha }\hfill %
\cr P_{ab}-\frac{1}{2}\Omega _{ab}P\hfill \cr
\end{matrix}
\right.
\end{equation}
where $P_{ab}$ is the $\frac{SO\left( 1,5\right) }{SO\left( 5\right) }$
vielbein, satisfying $P_{ab}\Omega ^{ab}=0$.

On the solution, the Hessian matrix (\ref{h6}) is then:
\begin{equation}
H_{ij}=2X^{2}\left( P^{ab}P_{ab}+3P^{2}\right) \,.
\end{equation}
As expected, the directions corresponding to the scalars in the $\mathbf{%
(4,2)}$ of $\frac{Sp(4,2)}{Sp(4)\times Sp(2)}$ are flat. When the theory is
interpreted as an $N=6$ one, this is the BPS solution whose states, regarded
as $N=2$ BPS multiplets, have flat directions corresponding to the
hyperscalar sector. On the other hand, in the $N=2$ interpretation this is
instead the non-BPS solution, and now the flat directions correspond to
degrees of freedom in the vector multiplets' moduli space.

The second solution (non-BPS $N=6$, BPS $N=2$) leaves all the $Sp(6)$
symmetry unbroken since the horizon value of the central charge matrix in
normal form is now:
\begin{equation}
Z_{AB}\rightarrow 0\,.
\end{equation}
Now the vielbein degrees of freedom do not decompose at all
\begin{eqnarray}
\mathbf{14} &\rightarrow &\mathbf{14}  \notag \\
P_{AB} &\rightarrow &P_{AB}
\end{eqnarray}
and correspondingly all the scalar degrees of freedom become massive.

Let us end this section by writing the quantities used here in the ${\ N}=2$
formalism adopted in subsubsection \ref{Subsubsect321}. In this case the
rigid index $a$ labelling the tangent space directions are replaced by the
antisymmetric traceless couple $[AB]$ (recall that we use the convention
that any summation over an antisymetrized couple always requires a factor $%
1/2$) :
\begin{eqnarray}
T_{A_{1}A_{2},B_{1}B_{2},C_{1}C_{2}} &=&2\,\sqrt{\frac{3}{2}}\,\left( \Omega
_{A_{1}B_{1}}\,\Omega _{B_{2}C_{1}}\,\Omega _{C_{2}A_{1}}-\frac{1}{6}%
\,\Omega _{A_{1}A_{2}}\,\Omega _{B_{1}C_{1}}\,\Omega _{B_{2}C_{2}}-\right.
\notag \\
&&\left. -\frac{1}{6}\,\Omega _{B_{1}B_{2}}\,\Omega _{A_{1}C_{1}}\,\Omega
_{A_{2}C_{2}}-\frac{1}{6}\,\Omega _{C_{1}C_{2}}\,\Omega
_{B_{1}A_{1}}\,\Omega _{B_{2}A_{2}}+\frac{1}{18}\,\Omega
_{A_{1}A_{2}}\,\Omega _{B_{1}B_{2}}\,\Omega _{C_{1}C_{2}}\right) \,;  \notag
\\
g_{A_{1}A_{2},B_{1}B_{2}} &=&\Omega _{B_{1}A_{1}}\,\Omega _{B_{2}A_{2}}-%
\frac{1}{6}\,\Omega _{A_{1}A_{2}}\,\Omega _{B_{1}B_{2}}\,,  \notag \\
&&
\end{eqnarray}
where antisymmetrization in the couples $(A_{1},A_{2}),\,(B_{1},B_{2}),%
\,(C_{1},C_{2})$ is understood. As far as the central charges are concerned,
we have the following correspondence:
\begin{equation}
Z=\frac{1}{\sqrt{3}}\,X\,\,;\,\,\,P^{a}\,Z_{a}=\frac{1}{2\,\sqrt{3}}%
\,P^{AB}\,Z_{AB}\,.
\end{equation}
\setcounter{equation}0
\def\theequation{4.\arabic{subsection}.\arabic{equation}}

\subsection{\label{N=4}$N=4$, $d=5$ and $(2,0)$, $d=6$}

The scalar manifold is the coset
\begin{equation}
G/H=O(1,1)\times \frac{SO(5,n)}{Sp(4)\times SO(n)}\,,  \label{n=4coset}
\end{equation}
spanned by the vielbein $d\sigma$, $P_{IAB}$ ($A,B=1,\cdots ,4$, $%
I=1,\cdots ,n$), where $d\sigma =\partial _{i}\sigma d\phi ^{i}$ is the
vielbein of the $O(1,1)$ factor while $P_{IAB}=P_{IAB,i}d\phi ^{i}$ is the $%
\Omega $-traceless vielbein of $\frac{SO(5,n)}{Sp(4)\times SO(n)}$
satisfying the conditions
\begin{equation}
P_{IAB}=P_{I[AB]}\,,\qquad P^{IAB}\Omega _{AB}=0\,.
\end{equation}
The bare electric charges are a $SO(5,n)$-singlet $e_0$ and a $SO(5,n)$-vector $e_\Lambda$ (the weight
with respect to $SO(1,1)$  is $+2$ for $e_0$ and $-1$ for $e_\Lambda$).

The BH potential reads:
\begin{equation}
V=\frac{1}{2}Z_{AB}Z^{AB}+4X^{2}+Z_{I}Z^{I}\,,  \label{pot4}
\end{equation}
and the differential relations among the $5$ central charges $Z_{AB}$
(satisfying $Z_{AB}\,\Omega ^{AB}=0$), the singlet $X$ and the $n$ matter
charges $Z_{I}$ are \cite{adf5d}:
\begin{eqnarray}
\nabla Z_{AB} &=&Z^{I}P_{IAB}-Z_{AB}d\sigma ; \\
\nabla X &=&2Xd\sigma ; \\
\nabla Z_{I} &=&\frac{1}{2}Z^{AB}P_{IAB}-Z_{I}d\sigma \,,  \label{diffrel4}
\end{eqnarray}
yielding
\begin{equation}
\nabla V=2P_{IAB}\left( Z^{AB}Z^{I}\right) +2d\sigma \left( 8X^{2}-\frac{1}{2%
}Z_{AB}Z^{AB}-Z_{I}Z^{I}\right) \,.
\end{equation}

The central charge matrix may be put in normal form:
\begin{equation}
Z_{AB}=
\begin{pmatrix}
e_{1} & 0\cr0 & -e_{1}
\end{pmatrix}
\otimes
\begin{pmatrix}
0 & 1\cr-1 & 0
\end{pmatrix}
\,,  \label{z4}
\end{equation}
so that the BH potential takes the form
\begin{equation}
V=2e_{1}^{2}+4X^{2}+Z_{I}Z^{I}\,,
\end{equation}
and the differential relations among the dressed charges become ($d\sigma $
and $P_{I}\equiv P_{I12}=-P_{I34}$ are the components of the scalar vielbein
coupling to the charges in normal form):
\begin{eqnarray}
\nabla e_{1} &=&Z^{I}P_{I}-e_{1}d\sigma \\
\nabla X &=&2Xd\sigma \\
\nabla Z_{I} &=&2e_{1}P_{I}-Z_{I}d\sigma \,.  \label{diffrel4nf}
\end{eqnarray}
Then the extremization of the BH potential takes the form
\begin{equation}
\nabla V=8P_{I}\left( e_{1}Z^{I}\right) +2d\sigma \left(
8X^{2}-2e_{1}^{2}-Z_{I}Z^{I}\right) =0\,.
\end{equation}

Two inequivalent solutions with finite area are there:

\begin{enumerate}
\item  $Z_{I}=0\,;\quad e_{1}=2X$. \newline
This is the $N=4$ 1/4-BPS solution and breaks the $Sp(4)$ $\mathcal{R}$%
-symmetry of the theory to $Sp(2)\times Sp(2)$, leaving the $SO(n)$ symmetry
unbroken. It corresponds to an $\frac{SO(5,n)}{SO(4,n)}$ orbit of the charge
vector.

\item  $Z_{AB}=0\,;\quad Z_IZ^I =8 X^2 $.\newline
It is a non-BPS attractor of the $N=4$ theory, corresponding to choose the
vector $Z_I$ to point in a given direction, say 1, in the space of charges: $%
Z_I= 2\sqrt 2 \delta_I^1$. This solution breaks the symmetry of the theory
to $Sp(4)\times SO(n-1)$, and corresponds to an $\frac{SO(5,n)}{SO(5,n-1)}$
orbit of the charge vector.
\end{enumerate}

In both cases the Bekenstein--Hawking entropy turns out to satisfy
\cite{adf5d}
\begin{equation}
S_{BH}^{(5)}=(V|_{extr})^{3/4}=\sqrt{|e_0 e^\Lambda e_\Lambda|}.
\end{equation}

To study the stability of the solutions, let us consider the Hessian
matrix
\begin{eqnarray}
H_{ij} &\equiv &\frac 14 \nabla _{i}\nabla _{j}V  \label{h4} \\
&=&P^{IAB}_{,i}P_{JCD,j}\left(\frac 14 Z_{AB}Z^{CD}\delta^I_{J}+\frac 12 Z^I
Z_J \delta_{AB}^{CD}\right) +  \notag \\
&&-2P^{IAB}_{,(i}\partial_{j)} \sigma Z_{AB} Z_I +\partial_i \sigma
\partial_j \sigma \left(\frac 12 Z_{AB}Z^{AB} + Z_I Z^I + 16 X^2\right)
\end{eqnarray}
and evaluate it on the two extrema.

On the BPS attractor solution, the $\mathcal{R}$-symmetry $Sp(4)$ is broken
to $Sp(2)\times Sp(2)$ ($A\rightarrow (\alpha \tilde{\alpha})$) and the
dressed charges in normal form become
\begin{equation}
Z_{AB}\rightarrow 2X
\begin{pmatrix}
\epsilon _{\alpha \beta } & 0\cr0 & -\epsilon _{\tilde{\alpha}\tilde{\beta}}
\end{pmatrix}
\,,\quad Z_{I}\rightarrow 0\,.
\end{equation}
Correspondingly, the vielbein $P^{IAB}$ decomposes to $(P^{I}\epsilon
^{\alpha \beta },-P^{I}\epsilon ^{\tilde{\alpha}\tilde{\beta}},P^{I\alpha
\tilde{\alpha}})$ where $P^{I}$ and $P^{I\alpha \tilde{\alpha}}$ are the
vielbein of the submanifold $\frac{SO(1,n)}{SO(n)}$ (spanning $N=2$ vector
multiplets) and $\frac{SO(4,n)}{Sp(2)\times Sp(2)\times SO(n)}$ (spanning $%
N=2$ hypermultiplets) respectively. Since on the solution $\frac{1}{2}%
Z_{AB}P^{IAB}\rightarrow 4XP^{I}$, the Hessian matrix (\ref{h4}) then
becomes:
\begin{equation}
H_{ij}=8X^{2}\left( 2P_{,i}^{I}P_{I,j}+3\partial _{i}\sigma \partial
_{j}\sigma \right)
\end{equation}
showing that the $4n$ scalars parametrized by $P^{I\alpha \tilde{\alpha}}$,
which correspond to $N=2$ hypermultiplets, have massless Hessian modes.

On the other hand, the non-BPS solution breaks the symmetry $SO(n) $ to $%
SO(n-1)$ ($I \to 1,k$; $k =1,\cdots n-1$) so that the vielbein $P_{IAB}$
decomposes into $(P_{1 AB}, P_{k AB})$. The Hessian matrix on the solution
is:
\begin{equation}
H_{ij}= 8X^2\left(\frac 12P^{1AB}_{,i}P_{1AB,j}+3\partial_i \sigma
\partial_j \sigma \right)
\end{equation}
Note in particular that the $5(n-1)$ scalars corresponding to the vielbein $%
P_{k AB}$, spanning the submanifold $\frac{SO(5,n-1)}{SO(5)\times SO(n-1)}$,
are flat directions.

For the $N=4$ theory it is easy to find a six dimensional uplift in
terms of the $IIB$, $(2,0)$ chiral $d=6$ theory coupled to $n$
tensor multiplets \cite{ADFL-6} (at least for the anomaly-free case
$n=21$) on similar lines as performed in section \ref{Sect3}.
Indeed, similarly to the dimensional reduction of the $N=2$ theory
coupled to tensor multiplets only, in the dimensional reduction of
the $IIB$ theory from six to five dimensions the
scalar content is incremented only by the KK-dilaton, which provides a $%
O(1,1)$ factor commuting with the $\frac{SO(5,n)}{SO(5)\times SO(n)}$ coset.
Moreover, the vector content in the gravitational multiplet is also
incremented by one graviphoton (whose integral corresponds to the singlet
charge $X$). Since the KK-dilaton is stabilized on the attractor solutions,
then the five dimensional attractors are in one to one correspondence with
the six dimensional ones: on the BPS attractor there are $4n$ flat
directions (corresponding to the quaternionic manifold $\frac{SO(4,n)}{%
SO(4)\times SO(n)}$), while on the non-BPS solution there are
$5(n-1)$ flat directions (spanning the coset $SO(5)\times
SO(n-1)$).\setcounter{equation}0
\def\theequation{\arabic{section}.\arabic{equation}}

\section{\label{Anomalies}Anomaly free $(1,0)$, $d=6$ supergravity with neutral matter}

In this section we comment on the constraints that an $N=2$, $d=5$
supergravity should satisfy in order to be uplifted to an an anomaly-free $%
N=\left( 1,0\right) $, $d=6$ theory.

It is well known that in a $\left( 1,0\right) $ supergravity with
neutral matter the absence of the gravitational anomaly demands a
relation among the triple $n_{T}$, $n_{V}$, $n_{H}$ of possible
matter multiplets (tensor, vector and hyper multiplets,
respectively), namely \cite{grav-free1, grav-free2}
\begin{equation}
n_{H}-n_{V}+29n_{T}=273.  \label{grav-anom-free}
\end{equation}

Moreover, the consistency of the gauge invariance of tensor and
(Abelian) vector multiplets requires that the gauge vector current
is conserved, \textit{i.e.}
\cite{FMS,AFL-6-5,gauge-free1,gauge-free2,gauge-free3}
\begin{equation}
d \,{{}^*J}_{\alpha }=\eta _{\Lambda \Sigma }C_{\alpha \beta
}^{\Lambda }C_{\gamma \delta }^{\Sigma }F^{\beta }\wedge F^{\gamma
}\wedge F^{\delta }=0,  \label{gauge-anom-free}
\end{equation}
implying that ($C_{\alpha \beta }^{\Lambda }=C_{(\alpha \beta )}^{\Lambda }$%
)
\begin{equation}
\eta _{\Lambda \Sigma }C_{(\alpha \beta }^{\Lambda }C_{\gamma \delta
)}^{\Sigma }=0.  \label{gauge-anom-free-2}
\end{equation}
Such a condition holds true for all symmetric real special manifolds
\cite{vanderseypen}, with the exception of the sequence $L\left(
-1,P\right) $, $P>0$ (whose corresponding K\"{a}hler and
quaternionic sequences are not symmetric \cite{dW-VP}). Disregarding
such a sequence, among all homogeneous real special spaces (see
\textit{e.g.} the Table 2 of \cite{vanderseypen}) the symmetric
spaces are $L\left( q,0\right) =L\left(
0,P\right) $, $q,P\mathbb{\geqslant }0$ (\textit{``}generic sequence\textit{%
''}, extended to consider also the $d=5$ uplift of the so-called $d=4$ $stu$
model), $L\left( q,1\right) $ for $q=1,2,4,8$ (magic supergravities over $%
J_{3}^{\mathbb{R}}$, $J_{3}^{\mathbb{C}}$, $J_{3}^{\mathbb{H}}$ and $J_{3}^{%
\mathbb{O}}$, respectively) and $L\left( -1,0\right) $ (the $d=5$ uplift of
the so-called $d=4$ $st^{2}$ model).

The condition (\ref{grav-anom-free}) for the magic models
respectively gives the following allowed triples $\left(
n_{T},n_{V},n_{H}\right) $ \cite {AFL-6-5}:
\begin{equation}
\begin{array}{l}
J_{3}^{\mathbb{R}}:\left( 2,2,217\right) ; \\
J_{3}^{\mathbb{C}}:\left( 3,4,190\right) ; \\
J_{3}^{\mathbb{H}}:\left( 5,8,136\right) ; \\
J_{3}^{\mathbb{O}}:\left( 9,16,28\right) .
\end{array}
\label{triplets}
\end{equation}
Notice that for the $J_{3}^{\mathbb{O}}$-based supergravity $n_{H}=28$, so
its corresponding quaternionic manifold could be identified with the
exceptional quaternionic K\"{a}hler coset \cite{CFG} $\frac{E_{8(-24)}}{%
E_{7}\times SU\left( 2\right) }$ (which is the quaternionic reduction - or
equivalently the hypermultiplets' scalar manifold - of the $d=4$ $J_{3}^{%
\mathbb{O}}$-based supergravity \cite{CFG,vanderseypen}).

On the other hand, for the \textit{``}generic sequence\textit{'' }there are
two possible uplifts to $d=6$, depending whether one starts with $L\left(
q,0\right) $ or $L\left( 0,P\right) $. Indeed, starting from $L\left(
q,0\right) $ the condition (\ref{grav-anom-free}) implies
\begin{equation}
n_{H}=244-29q,
\end{equation}
which demands $0\leqslant q\leqslant 8$, whereas starting from $L\left(
0,P\right) $ the same anomaly-free condition yields
\begin{equation}
n_{H}=244+P,
\end{equation}
which always admits a solution.

The $(1,0)$, $d=6$ theory obtained by uplifting the real special
symmetric sequence $L\left( q,0\right) $ has $n_{V}=0$ and
$n_{T}=q+1$, and thus $1\leqslant n_{T}\leqslant 9$. On the other
hand, the anomaly-free $(1,0)$, $d=6$ uplift of the real special
symmetric sequence $L\left( 0,P\right) $ has $n_{T}=1$ and $n_{V} $
arbitrary, thus it may be obtained from the standard
compactification of heterotic superstrings on $K_{3}$ manifolds (see
\textit{e.g.} \cite{GSW}).

The model $L\left( -1,0\right) $ admits an anomaly-free uplift to $d=6$,
having $n_{V}=n_{T}=0$ and $n_{H}=273$.

All other homogeneous non-symmetric real special spaces do not
fulfill the condition
(\ref{gauge-anom-free})-(\ref{gauge-anom-free-2}) in presence of
only neutral matter, so they seemingly have a $d=6$ uplift to
$(1,0)$ supergravity which is not anomaly-free, unless they are
embedded in a model where a non-trivial gauge group is present, with
charged matter \cite{open,anto}.

\section{\label{Conclusion}Conclusion}

There are three theories with eight supercharges which admit black
hole/black string attractors, namely $N=2$ supergravity in $d=4$,
$5$, $6$ dimensions. For symmetric special geometries, the entropy
is respectively given by the quartic, cubic and quadratic invariant
of the corresponding $U$-duality group in the three diverse
dimensions. In this paper we extend previous work \cite{CFM1} on the
investigation of the BPS and non-BPS attractor equations of such
theories, by relating them as well as the corresponding moduli
spaces of (non-BPS) critical points.

Furthermore, we related the moduli space of the $N=8$, $d=5$ BPS
unique orbit to the moduli space of $N=2$, $d=5$ non-BPS orbit for
all magic supergravities, as well as for the \textit{``}generic
sequence\textit{''} of real special symmetric spaces. This latter is
directly related to the $d=6$ tensor multiplets' non-BPS moduli
space, which describes a neutral\textit{\ }dyonic superstring in
$d=6$.

We also considered $N=4$, $d=5$ supergravity, and related its $\frac{1}{4}$%
-BPS and non-BPS attractors to the ones of $(2,0)$, $d=6$ theory.
Also in this case the moduli space of non-BPS attractors is spanned
by the $d=6$ non-BPS flat directions, studied in \cite{ADFL-6}.

We stress that our analysis is purely classical and it does not deal
with quantum corrections to the entropy, so it should apply only to
the so-called ``large'' black objects. We leave the study of the
quantum regime to future work.

\section*{\textbf{Acknowledgments}}

We would like to acknowledge enlightening discussions with Massimo
Bianchi and Augusto Sagnotti.

A. M. would like to thank the Department of Physics, Theory Unit
Group at CERN for its kind hospitality and support during the
completion of the present paper.

The work of L.A., S.F. and M.T. has been supported in part by European
Community Human Potential Program under contract MRTN-CT-2004-005104 \textit{%
``Constituents, fundamental forces and symmetries of the
universe'',} in which L.A. and M.T. are associated to Torino
University, and S.F. is associated to INFN Frascati National
Laboratories. The work of S.F. has also been supported in part by
D.O.E. grant DE-FG03-91ER40662, Task C.

\setcounter{equation}0
\def\theequation{A.\arabic{equation}}

\section*{\label{Sect5}Appendix : Relevant Embeddings}

Let us first fix the notations to be used in the this appendix. If
$\alpha$ is a root of a complex Lie algebra $\frak{g}$, the
normalizations of the corresponding non-compact Cartan generator
$H_\alpha$ and of the shift generators $E_{\pm\alpha}$ will be
defined as follows \cite{slansky}:
\begin{eqnarray}
H_\alpha &=&\frac{2}{(\alpha\cdot\alpha)}\,\alpha^i\,H_i\,\,;\,\,\,\,(H_i,%
\,H_j)=\delta_{ij}\,,  \notag \\
E_{-\alpha}&=&(E_\alpha)^\dagger\,\,;\,\,\,\,(E_\alpha,\,E_{-\alpha})=\frac{2%
}{(\alpha\cdot\alpha)}\,,
\end{eqnarray}
where $(\cdot,\cdot)$ is the Killing form. The above normalizations imply
the following commutation relations
\begin{eqnarray}
[H_\alpha,\,E_\beta]&=&\langle\beta,\,\alpha\rangle\,E_\beta\,\,;\,\,\,[E_%
\alpha,\,E_{-\alpha}]=H_\alpha\,,  \notag \\
\langle\beta,\,\alpha\rangle&=&\frac{2}{(\alpha\cdot\alpha)}%
\,\beta\cdot\alpha\,.
\end{eqnarray}

\subsection*{\label{Subsect52}$J_{3}^{\mathbb{C}},d=5$ : the $\mathrm{SU}%
(2,1)^{2}\subset \mathrm{F}_{4(4)}$ embedding}

The simple roots of the $\frak{sl }(3,\mathbb{C})^2$ subalgebra of $\frak{f}%
_{4}$ over $\mathbb{C}$ are defined in terms of the simple roots of the
latter $\alpha_k$ ($k=1,\dots,4$, $\alpha_1,\,\alpha_2$ being long roots) as
follows
\begin{eqnarray}
a_1&=&\alpha_4\,\,,\,\,\,a_2=\alpha_3\,\,,\,\,\,b_1=\alpha_1\,\,,\,\,\,b_2=%
\alpha_1+3\,\alpha_2+4\,\alpha_3+2\,\alpha_4\,.
\end{eqnarray}
The real form $\frak{f}_{4(4)}$ contains an $\frak{sl}(2,\,\mathbb{R})^4$
subalgebra defined by the following mutually orthogonal roots:
\begin{eqnarray}
a_2\,\,,\,\,\,\,b_2\,\,,\,\,\,\,c=\alpha_1+\alpha_2+\alpha_3\,\,,\,\,\,\,d=%
\alpha_1+\alpha_2+2\,\alpha_3+2\,\alpha_4\,.
\end{eqnarray}
We can define the roots of $\frak{f}_{4(4)}$ using a Cartan subalgebra $%
\frak{h}_0$ generated by two non-compact $H_{a_2},\,H_{b_2}$ and two compact
$\mathrm{i}\,H_c,\,\mathrm{i}\,H_d$ generators, the latter corresponding to
the $\frak{so}(2)$ generators inside $\frak{sl}(2,\,\mathbb{R})_c\oplus
\frak{sl}(2,\,\mathbb{R})_d$. In terms of the generators of $\frak{h}_0$, we
can choose a basis of Cartan generators for $\frak{sl }(3,\mathbb{C})^2$ to
consist of $H_{a_2},\,H_{b_2}$ as well as of
\begin{eqnarray}
H_{a_1}&=&-\frac{1}{2}\,\left(H_{a_2}+\mathrm{i}\,H_{c}-2\,\mathrm{i}%
\,H_d\right)\,\,,\,\,\,H_{b_1}=-\frac{1}{2}\,\left(H_{a_2}-\mathrm{i}\,H_{c}-%
\mathrm{i}\,H_d\right)\,.
\end{eqnarray}
These generators define the Cartan subalgebra of an $\frak{su}(2,1)^2$
subalgebra of $\frak{f}_{4(4)}$. Indeed one can verify that the $\frak{sl }%
(3,\mathbb{C})^2$ root system defined by the simultaneous eigenvalues of the
$\frak{h}_0$ generators, is stable with respect to the conjugation $\sigma$
relative to $\frak{f}_{4(4)}$, namely that
\begin{eqnarray}
a_2^\sigma &=& a_2\,;\,\,\,a_1^\sigma=-(a_1+a_2)\,\,;\,\,\,b_2^\sigma =
b_2\,;\,\,\,b_1^\sigma=-(b_1+b_2)\,.
\end{eqnarray}
The $\frak{su}(2,1)^2$ generators are thus defined by $\sigma$--invariant
combinations of the $\frak{sl }(3,\mathbb{C})^2$ shift generators. The fact
that this construction defines an $\frak{su}(2,1)^2$ subalgebra of $\frak{f}%
_{4(4)}$ and not an $\frak{sl}(3,\mathbb{R})^2$ algebra is proven by the
existence in each factor of a compact Cartan subalgebra, defined by the
generators $\{E_{a_2}-E_{-a_2},\,\mathrm{i}\,(H_{c}- 2\,H_d)\}$ for the
first factor and $\{E_{b_2}-E_{-b_2},\,\mathrm{i}\,(H_{c}+ H_d)\}$ for the
second.

\subsection*{\label{Subsect54}$J_{3}^{\mathbb{R}},d=5$ : the $\mathrm{SL}(2,%
\mathbb{R})\times G_{2(2)}\subset \mathrm{F}_{4(4)}$ embedding}

Denoting by $a$ the $\frak{sl}(2,\mathbb{R})$ root and by $b_1,\,b_2$ the
simple roots of $\frak{g}_{2(2)}$, the $\mathrm{SL}(2,\mathbb{R})\times
G_{2(2)}$ generators can be written in terms of the $\mathrm{F}_{4(4)}$
generators as follows:
\begin{eqnarray}
H_{b_1}&=&H_{\alpha_1+\alpha_2}+H_{\alpha_4}\,\,;\,\,\,H_{b_2}=H_{\alpha_2+2%
\alpha_3}=H_{\alpha_2}+H_{\alpha_3}\,\,;\,\,\,
E_{b_1}=E_{\alpha_1+\alpha_2}+E_{\alpha_4}\,,  \notag \\
E_{b_2}&=&E_{\alpha_2+2\alpha_3}\,\,;\,\,\,E_{b_1+b_2}=-E_{\alpha_1+2%
\alpha_2+2\alpha_3}+E_{\alpha_2+2\alpha_3+\alpha_4} \,,  \notag \\
E_{2b_1+b_2}&=&-E_{\alpha_1+2\alpha_2+2\alpha_3+\alpha_4}+E_{\alpha_2+2%
\alpha_3+2\alpha_4}\,\,;\,\,\,
E_{3b_1+b_2}=-E_{\alpha_1+2\alpha_2+2\alpha_3+2\alpha_4}\,,  \notag \\
E_{3b_1+2b_2}&=&E_{\alpha_1+3\alpha_2+4\alpha_3+2\alpha_4}\,,  \notag \\
H_a&=&2\,(H_{\alpha_3+\alpha_4}+H_{\alpha_1+\alpha_2+\alpha_3})\,\,;\,\,%
\,E_a=\sqrt{2}\,(E_{\alpha_3+\alpha_4}+E_{\alpha_1+\alpha_2+\alpha_3})\,.
\end{eqnarray}

\subsection*{Matrix representation of $\frak{f}_{4}$ generators}

For the sake of completeness, let us give below an explicit realization of
the generators $H_{\alpha_i},\,E_{\alpha_i}$ and $\frak{f}_{4}$, in the
fundamental representation.

\paragraph{$\frak{f}_{4}$ generators:}

\begin{eqnarray}
H_{\alpha_1}&=&\mathrm{diag}%
(-1,1,0,0,0,0,0,0,0,-1,0,0,0,-1,1,-1,-1,1,1,1,0,-1,1, 0,0,0,0)\,,  \notag \\
H_{\alpha_2}&=&\mathrm{diag}(0, 0, 1, -1, 0, 0, 0, 1, 1, -1, -1, 0, 0, 0, 0,
1, 1, -1, -1, 0, 0, 0, 1, -1, 0, 0)\,,  \notag \\
H_{\alpha_3}&=&\mathrm{diag}(0, 1, -1, 1, -1, 1, 0, -1, 0, 1, 2, -1, 0, 0,
1, -2, -1, 0, 1, 0, -1, 1, -1, 1, -1, 0)\,,  \notag \\
H_{\alpha_4}&=&\mathrm{diag}(1, -1, 0, 0, 1, 0, -1, 1, -1, 1, -1, 2, 0, 0,
-2, 1, -1, 1, -1, 1, 0, -1, 0, 0, 1, -1)\,,  \notag \\
E_{\alpha_1}&=&I_{4,6}+I_{5,8}+I_{7,9}+I_{18,20}+I_{19,22}+I_{21,23}\,,
\notag \\
E_{\alpha_2}&=&I_{3,4}+I_{8,10}+I_{9,11}+I_{16,18}+I_{17,19}+I_{23,24}\,,
\notag \\
E_{\alpha_3}&=&I_{2,3}+I_{4,5}+I_{6,8}+I_{10,12}+c_1\,I_{11,13}+c_2%
\,I_{11,14}+c_1\,I_{13,16}+c_2\,I_{14,16}+I_{15,17}+I_{19,21}+  \notag \\
&&+ I_{22,23}+I_{24,25}\,,  \notag \\
E_{\alpha_4}&=&I_{1,2}-I_{5,7}-I_{8,9}-I_{10,11}+c_2\,I_{12,13}+c_1%
\,I_{12,14}+c_2\,I_{13,15}+c_1\,I_{14,15}-I_{16,17}-I_{18,19}-  \notag \\
&&- I_{20,22}+I_{25,26}\,,
\end{eqnarray}
where $c_1=(1+\sqrt{3})/2$, $c_2=(1-\sqrt{3})/2$ and $(I_{I,J})_{KL}=%
\delta_{IK}\,\delta_{JL}$. The Killing form is $(M_1,M_2)=\frac{1}{6}\,%
\mathrm{Tr}(M_1\,M_2)$.

\end{document}